\documentclass[fleqn,usenatbib]{mnras}

\usepackage{newtxtext,newtxmath}
\usepackage[T1]{fontenc}
\usepackage{ae,aecompl}
\usepackage{graphicx}    
\usepackage{amsmath}      
\usepackage{siunitx}
\usepackage{IEEEtrantools}

\def\UMi {{$\beta$~UMi~b~}}
\def\alfT {{$\alpha$~Tau~b~}}

\title[A uGMRT search for radio emission from planets around  evolved stars]{A uGMRT search for radio emission from planets around  evolved stars}

\author[Mayank Narang]{
Mayank Narang$^{1,2}$,\thanks{E-mail: mnarang@asiaa.sinica.edu.tw}
P. Manoj$^{2},$
C. H. Ishwara Chandra$^{3}$,  Bihan  Banerjee$^{2}$, \newauthor Himanshu Tyagi$^{2}$ , Motohide Tamura$^{4,5,6}$, Thomas Henning$^7$, Blesson Mathew$^8$
\newauthor Joseph Lazio$^9$ ,  Arun Surya $^{2}$, Prasanta K. Nayak $^{10,2}$
\\
$^{1}$ Academia Sinica Institute of Astronomy \& Astrophysics, 11F of Astro-Math Bldg., No.1, Sec. 4, Roosevelt Rd., Taipei 10617, Taiwan, R.O.C.\\
$^{2}$Tata Institute of Fundamental Research 
Homi Bhabha Road, Colaba, Mumbai 400005, India\\
$^{3}$National Centre for Radio Astrophysics, TIFR, Post Bag 3, Ganeshkhind, Pune 411007, India\\ 
$^4$ The University of Tokyo, 7-3-1, Hongo, Bunkyo-ku, Tokyo, 113-0033, Japan\\
$^5$ Astrobiology Center, 2-21-1, Osawa, Mitaka, Tokyo 181-8588, Japan\\
$^6$ National Astronomical Observatory of Japan, 2-21-1 Osawa, Mitaka, Tokyo 181-8588, Japan\\ 
$^7$ Max-Planck-Institut f\"{u}r Astronomie, K\"{o}nigstuhl 17, D-69117 Heidelberg, Germany\\
$^{8}$Department of Physics and Electronics, CHRIST (Deemed to be University), Bangalore, Karnataka - 560034\\
$^{9}$ Jet Propulsion Laboratory, California Institute of Technology, Pasadena CA 91106, United States\\
$^{10}$ Institute of Astrophysics, Pontificia Universidad Católica de Chile, Av. Vicuña MacKenna 4860, 7820436, Santiago, Chile
}

\begin{document}
\label{firstpage}
\pagerange{\pageref{firstpage}--\pageref{lastpage}}
\maketitle

\begin{abstract}
In this work, we present the results from a study using the  Giant Meterwave Radio Telescope (GMRT) to search for radio {emission} from planets around three evolved stars namely $\alpha$~Tau, $\beta$~UMi, and $\beta$~Gem. Both  $\alpha$~Tau and $\beta$~UMi host massive $\sim$ 6 $M_J$ mass planets at about $\sim$1.4 au from the central star, while $\beta$~Gem is host to a 2.9 $M_J$ mass planet at 1.7 au from the host star. We observe $\alpha$~Tau and $\beta$~ UMi at two u(upgraded)GMRT bands; band~3 (250-500~MHz) and band~4 (550-900~MHz). We also analyzed the archival observations from $\beta$ Gem at 150~MHz from GMRT. We did not detect any radio signals from these systems. At 400~MHz, the 3$\sigma$ upper limit is 87 $\mu$Jy/beam for $\alpha$~Tau~{b} and 77.4 $\mu$Jy/beam for $\beta$~UMi~{b}. From our observations at 650~MHz, we place a 3$\sigma$ upper limit of 28.2 $\mu$Jy/beam for $\alpha$~Tau~b and 33.6 $\mu$Jy/beam for $\beta$~UMi~b. For $\beta$ Gem b, at 150~MHz, we place an upper limit of 2.5 mJy. At 400~MHz and 650~MHz, our observations are the deepest radio images for any exoplanetary system.  

\end{abstract}


\section{Introduction}

\begin{figure*}
\centering
\includegraphics[width=1\linewidth]{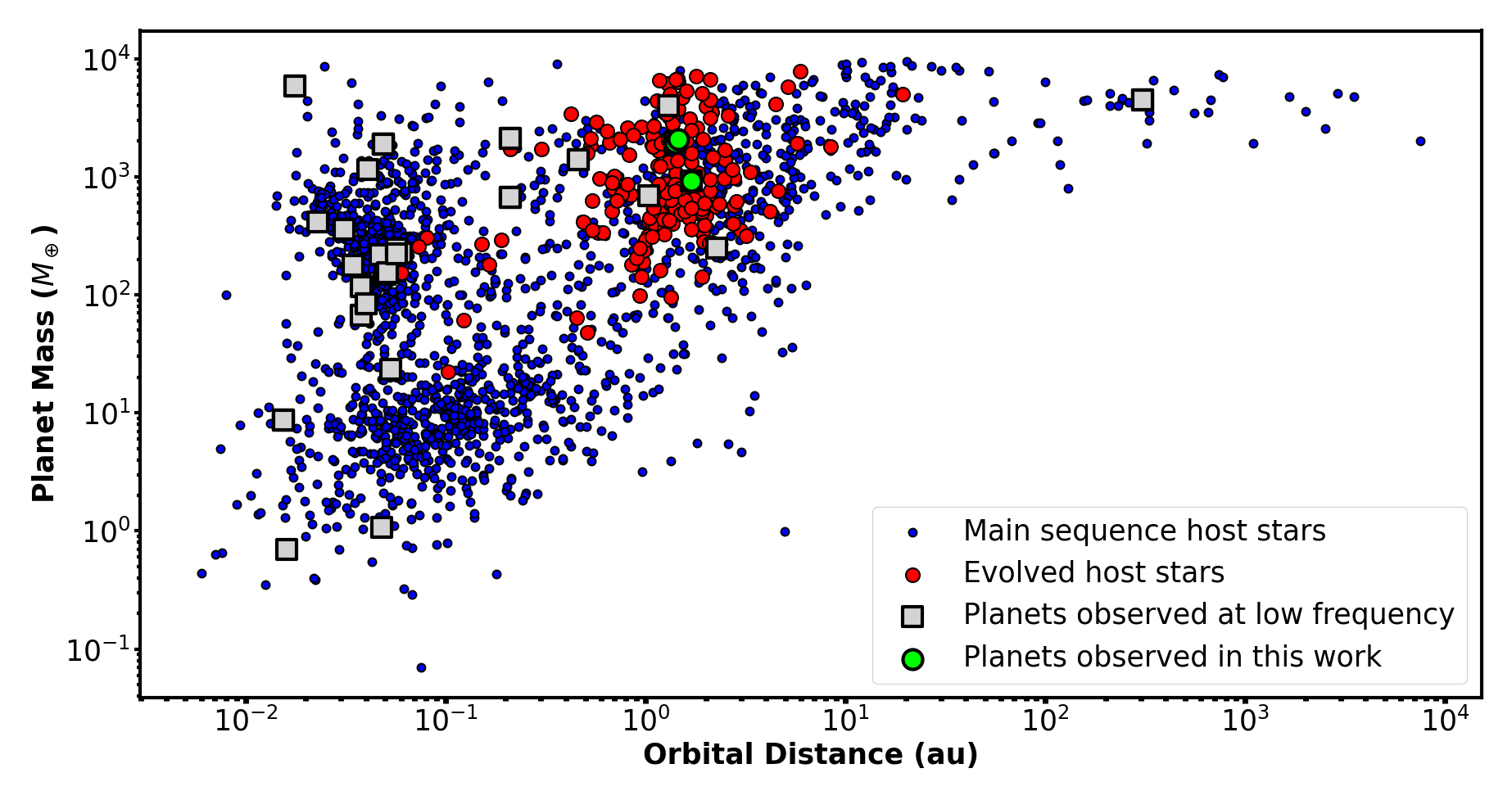}
\caption{{The orbital distance and mass distribution of all exoplanet systems that have a targeted observation at low frequency ($<$ 2 GHz) are shown as grey squares. These grey squares are overplotted on the current distribution of all known exoplanets (retrieved on 2023 September 11). The planets around main sequence stars are shown in blue solid circles while the planets around evolved stars are shown as red solid circles.  We also mark the position of $\alpha$~Tau~b, $\beta$~UMi~b, and $\beta$~Gem~b, the three exoplanets that we study in this work as green circles. }}
\label{Dem}
\end{figure*}

Over the past three decades, the field of exoplanet research has progressed significantly. It has transitioned from basic detections of exoplanets to a more comprehensive analysis, including characterizing individual planetary systems and studying exoplanets as a population 
. One of the emerging frontiers in exoplanet exploration is the investigation of their magnetic fields. The radio {emission} originating from exoplanets can provide unambiguous evidence of these magnetic fields \citep{Griessmeier15}. This, in turn, can provide valuable information about the internal composition and internal dynamics of these exoplanets \citep{Sanchez-Lavega04,  Reiners10, Lazio19}. {Despite extensive efforts, only one tentative detection of radio emission from an exoplanet \citep{Turner20} has been reported so far.} In Figure \ref{Dem}, we show the orbital distance and mass distribution of all exoplanet systems that have a targeted observation at low frequency ($<$ 2 GHz). For systems with multiple exoplanets, we only show the innermost planet. {We can see most studies on radio {emission} from exoplanets have primarily focused on hot Jupiters, which are gas giants ($M_P>0.3 M_J$) with orbital distances $\leq$ 0.3 au. The parameter space of long-period warm and cold Jupiters ($M_P>0.3 M_J$ and orbital distance >0.3 au) has not been extensively probed.  }

Recent studies \citep[e.g.,][]{gorman18,Lynch18, Narang21} have argued that hot Jupiters may not be the most suitable candidates for observing radio {emission} \citep[also see ][]{Jardine08, Nichols16, Daley17, Weber17}. Hot Jupiters are believed to be tidally locked, resulting in low rotation rates that lead to weakened magnetic fields and, subsequently, weaker radio {emission}. Additionally, since these hot Jupiters are located very close to their host stars, they may be situated within the host star's magnetosphere. The magnetospheres of such close-in planets are then determined by the magnetic pressure exerted by the stellar magnetosphere \citep{Jardine08}. As the distance between the star and the planet decreases, the planet's magnetosphere contracts while the electron density (which is a proxy for the input power supplied to the planet for radio emission) increases. This increase in electron density is counteracted by the shrinking magnetosphere (due to the pressure from the stellar magnetosphere), leading to the radio emission output power saturating as the input power increases  \citep{Jardine08}.


Moreover, in the case of hot Jupiters, the magnetospheric convection may also saturate, resulting in these planets being unable to dissipate the incident magnetic energy from the stellar wind \citep{Nichols16}. This could cause the expected flux density of radio emission from hot Jupiters to be overestimated. For the radio emission from the exoplanets to propagate, the local plasma frequency must be less than the emission frequency. For close-in Jupiters, the outer layers of the atmosphere are highly ionized, due to which the upper atmosphere can have a high plasma frequency and hence prevent emission from propagating \citep[e.g.,][]{Daley17, Weber17,  Weber18}. Based on these factors, hot Jupiters around main sequence stars might not be the ideal candidates to observe at radio wavelengths. However, most of the shortfalls of observing hot Jupiters around main-sequence stars can be overcome by observing planets around evolved stars.


Evolved stars are stars that have evolved off the main sequence and are no longer burning hydrogen in their core \citep{1996ima..book.....C}.  Most of the exoplanets that have been detected around evolved or post-main-sequence host stars are Red Giant Branch (RGB) stars. The RGB marks the phase of stellar evolution that precedes the helium burning in the core and is marked by hydrogen fusion in the shell around the core \citep{1996ima..book.....C}.   The system architecture of planets around evolved stars is very different from their main sequence counterparts. As can be seen from Figure \ref{Dem}, planets that orbit evolved stars tend to be massive Jupiter-like planets ($M_P\ga 1 M_J$) \citep[e.g.][]{Lovis07, Dollinger09, Jones14} that orbit at distances $\ga$1.0  au  \citep[e.g.][]{Johnson07, Sato08, Kunitomo11, Jones14}. 

Since these planets are not very close to their host stars, they are not tidally locked. Due to this, they are likely to have higher rotation rates. The higher rotation rates coupled with these planets being massive suggests that they might possess stronger magnetic fields than hot Jupiters. The stellar mass-loss rates from the {RGB stars} are  4 -- 5 orders of magnitude higher than that for the main sequence stars \citep{Reimers75, 2005ApJ...630L..73S, Cranmer11,Gorman13,McDonald15}. The radio emission from exoplanets is proportional to the  {ionized} mass-loss rate \citep[e.g.,][]{Zarka01a, gorman18}. Even though planets around evolved stars orbit at larger distances ($\geq$ 1 au), the host star's enhanced mass-loss rate (by 4--5 orders) compensates for the larger orbital distances and can lead to the generation of radio emission comparable or higher in strength to their main sequence hot-Jupiter counterparts. Hence giant planets around evolved stars might serve as better targets to search for radio emission due to star-planet interaction. Furthermore, radio observations of only three planetary systems ($\beta$ Gem, $\iota$ Dra, and $\beta$ UMi) around evolved stars have been carried out \citep{gorman18}. These observations were carried out with LOw Frequency ARray (LOFAR) at 150 MHz but did not result in any detections. Since the radio emission from planets around evolved stars has not been extensively studied, this provides an excellent and novel parameter space.  

This work presents results from our upgraded Giant Meterwave Radio Telescope (uGMRT) observations in band~3 (250-500~MHz) and band~4 (550-900~MHz) for two planet-hosting evolved stars, $\alpha$~Tau, and $\beta$ UMi. Additionally, we have analyzed the archival observations for the system $\beta$ Gem at 150~MHz with GMRT. These three exoplanets show the largest expected flux densities (based on our calculations in Section 2) among the known exoplanets around evolved stars and have their maximum cyclotron frequencies between 150 - 650~MHz. Hence these are ideal targets to be observed at low radio frequencies with uGMRT. In Section 2, we discuss our sample selection method and the targets, while in  Section 3, we describe the observational setup and the data reduction steps. Next, we present and discuss our results in Sections 4 and 5, and finally, we summarize in Section 6. 

\section{Sample selection}
To select evolved stars among the known planet-hosting stars, we used the criteria followed by \cite{Ciardi2011} and \cite{Gayathri20}. An evolved star is one that is in the $\mathrm{log~\textsl{g}}$ range:

\begin{equation}
\mathrm{log~{g}} < \left\{ \,
\begin{IEEEeqnarraybox}[][c]{l?s}
\IEEEstrut
3.5 & if ${T_\mathrm{eff} \,\mathrm{(K)}} \geq 6000$ \\
4.0 & if ${T_\mathrm{eff} \, \mathrm{(K)}} \leq 4250$ \\
5.2 - (2.8 \times 10^{-4}\,{T_\mathrm{eff}}) & if $4250<  {T_\mathrm{eff} \, \mathrm{(K)}}< 6000$.
\IEEEstrut
\end{IEEEeqnarraybox}
\right.
\label{MSeq}
\end{equation}

Using Equation \ref{MSeq}, we classified all the host stars from the table of the confirmed planets at the NASA Exoplanet Archive \citep{akeson13} as either main-sequence stars or evolved stars. Our initial sample consisted of 4009 planets around 2980 host stars (retrieved on July 4, 2019). Out of these 2980 host stars, we found 168 host stars that have evolved off the main sequence. These evolved stars host 187 planets around them, with 19 systems hosting multiple planets. For this study, we selected planetary systems with predicted flux densities {$\geq$} 0.1~mJy and a maximum cyclotron frequency of $\nu_c$ falling in either band~3 (250-500~MHz) or band~4 (550-900~MHz) of uGMRT (see below). This frequency range was selected because uGMRT is the most sensitive at these frequencies \citep{YG}. The flux density limit was chosen to ensure at least a $3\sigma$ detection at these bands. The expected rms with an integration time of $\sim$ 4 hrs in band~3 (250-500~MHz) and band~4 (550-900~MHz) of uGMRT is about 20-30 $\mu$Jy/beam. 

For our flux density estimates, we used the kinetic model similar to those of \cite{Ignace09} and \cite{gorman18} {(see Equation \ref{flux_er})}. To estimate the flux density from an exoplanet around an evolved star, we first calculate the maximum cyclotron frequency $\nu_c$ of emission from the exoplanet and the mass loss rate from the host star $\dot M$. The  maximum  cyclotron frequency we computed using the formalism from \cite{farrell99}:

\begin{equation}
    \nu_{c}=23.5 \times \frac{\omega}{\omega_J}\times \left(\frac{M_P}{M_J}\right)^{5/3}  \times \left(\frac{R_J}{R_P}\right)^{3} \mathrm{MHz}
    \label{nuer}
\end{equation}
where $\omega$ is the rotation rate of the planet, $\omega_J$ is the rotation rate of Jupiter, $M_P$ is the mass of the planet, $M_J$ is the mass of Jupiter, and $R_P$ is the radius of the planet, and $R_J$ is the radius of Jupiter.

The mass loss from the star $\dot{M_*}$,  can be calculated in terms of the stellar luminosity $L_*$, stellar radius $R_*$, {stellar effect temperature $T_\mathrm{eff}$} and the surface gravity $\textsl{g}$ from the star is given as \citep[][]{2005ApJ...630L..73S}:

\begin{equation}
  \dot{M_*} = \textstyle 8 \times 10^{-14} \times   \left(\frac{L_{\mathrm{bol}}}{\textsl{g}R_{*}}\right) \\ \left(\frac{T_\mathrm{eff}}{4000~\mathrm{K}}\right)^{3.5} \left(1+\frac{g_{\odot}}{4300\times g}\right)  M_\odot /\mathrm{yr}
\label{mass_losseq}
\end{equation}

{In the above express the entire stellar wind is assumed to be ionized, but \cite{1986AJ.....91..602D} have shown that for stars cooler than spectral type $\sim$K1 the stellar wind is only partially ionized. Therefore in this work, we assume that an ionized mass loss rate $\dot{M}_{ion}$ = 0.2 $\times \dot{M_*}$ for cooler K giants and $\dot{M}_{ion}$ =  $\dot{M_*}$ for hotter giants. }

 \begin{figure*}
\centering
\includegraphics[width=1\linewidth]{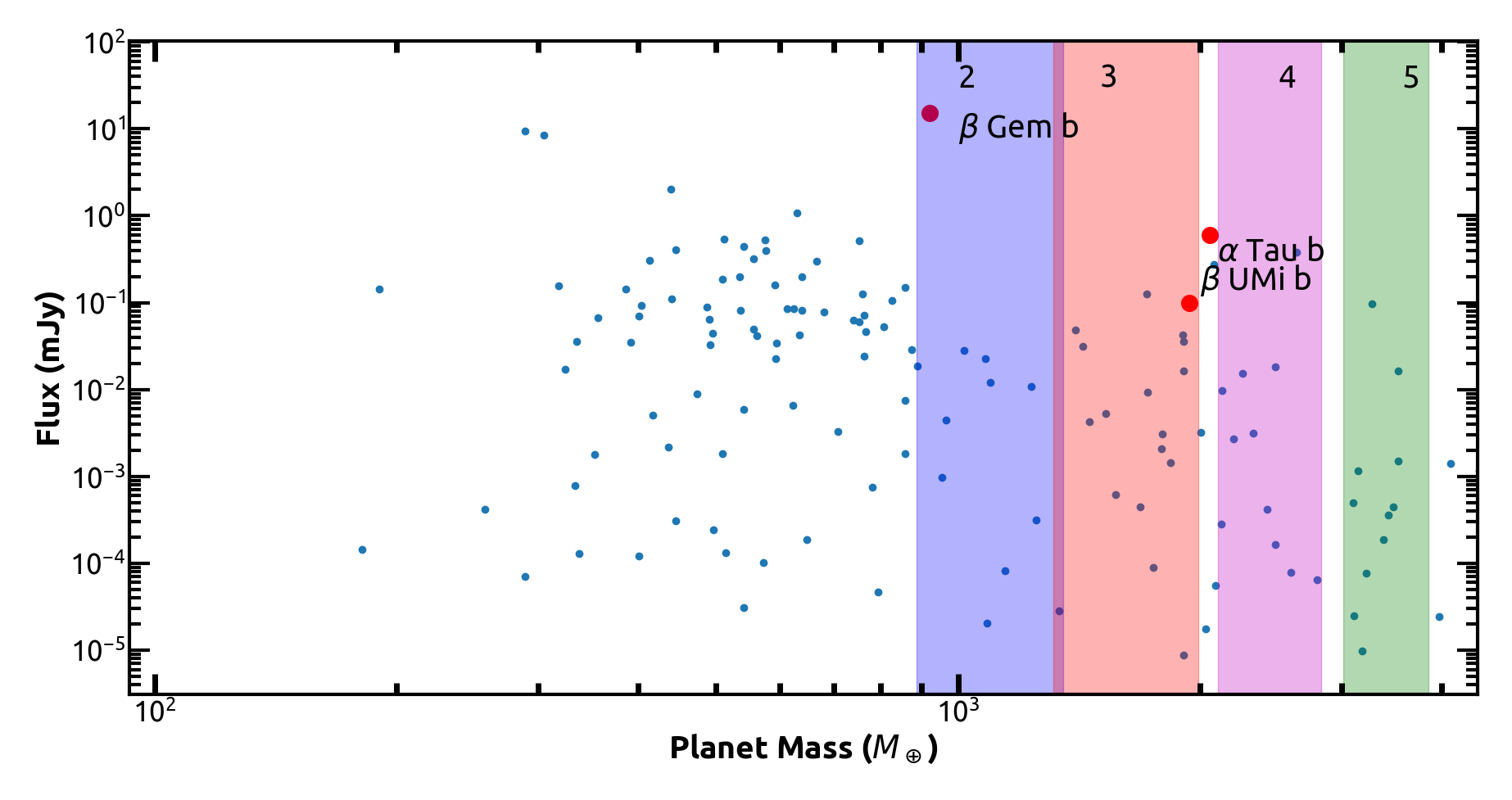}
\caption{The modeled flux density for all the planets around evolved stars from the  NASA Exoplanet Archive as a function of planet mass. Also shown are our target exoplanets (solid red circles) with their modeled flux densities. The uGMRT bands are also highlighted as colored bands in the plot. The bands shown are obtained by converting the emission frequency to the planet's mass using Equation \ref{flux_er} and assuming a rotation rate for the planets similar to that of Jupiter.  }
\label{fig1er}
\end{figure*}

The estimated flux density $S_\nu$ from the exoplanet can then be calculated as \citep{Ignace09,gorman18}:

\begin{equation}
\begin{aligned}
S_{\nu} =  4.6\mathrm{mJy}\,\times \left(\frac{\omega}{\omega_J}\right)^{-0.2} \, \left(\frac{M_P}{M_J}\right)^{-0.33} \, \left(\frac{R_P}{R_J}\right)^{-3} 
 \left(\frac{\Omega}{1.6 sr}\right)^{-1}\, \\   \times \left(\frac{d}{10\, \mathrm{pc}}\right)^{-2}\,   \left(\frac{a}{1 \, au}\right)^{-1.6} 
\left(\frac{\dot{M}_{ion}}{10^{-11} M_\odot  \mathrm{yr}^{-1} }\right)^{0.8} \left(\frac{V_\infty}{100 \mathrm{km} \mathrm{s}^{-1}}\right)^{2} 
\label{flux_er}
\end{aligned}
\end{equation}
where {$\dot{M}_{ion}$ is the ionized mass loss rate from the star},  $d$ is the distance to the exoplanet, $a$ is the semi-major axis, $V_\infty$ is stellar wind speed, and $\Omega$ is the solid angle of emission which we fix to 1.6 $sr$ \citep[similar to ][]{gorman18}. {For fully ionized stellar wind (from host stars hotter than K1) we assume $V_\infty$  = $0.75\times v_{esc}$, where $v_{esc}$ is the photospheric escape velocity \citep{1986AJ.....91..602D} and for partially ionized winds we assume that $V_\infty$ = 30 km/s. We further assume that all planets have a planetary radius $R_P$ = $R_J$} (since most planets detected around evolved stars are detected with the RV method and lack radius measurements) and rotation rate $\omega_J$, similar to that of Jupiter. These assumptions were made to ensure a homogeneous derivation of the radio flux density.

Using Equation \ref{nuer} and Equation \ref{flux_er}, we can calculate the maximum cyclotron frequency and expected flux density for all exoplanets around evolved stars. Figure \ref{fig1er} shows the modeled flux densities for all the planets orbiting evolved stars as a function of planet mass. We have further used Equation \ref{nuer} to convert the uGMRT frequency bands into a mass range (assuming the planets have a rotation rate of 10 hrs and a radius similar to that of Jupiter), which are highlighted as colored numbered bands in Figure \ref{fig1er}.

Five exoplanets fulfilled the criteria we had set, i.e., (i) maximum cyclotron frequency in band-3 and band-4 of uGMRT, and (ii) the expected flux density from the planets should be greater than 0.1 mJy. {For this study, we selected the two sources: which are $\beta$~ UMi~b and $\alpha$~Tau~b} (see Table \ref{Table1er} and Figure \ref{fig1er}). We also retrieved the archival GMRT data of the $\beta$ Gem planetary system.  In Figure \ref{Dem}, we show the three systems in the context of all previous exoplanet systems that have been observed at low frequency ($<2$ GHz). The three systems are described in greater detail in the following subsections.

\begin{table*}
\begin{center}
\caption{The exoplanet sample for GMRT observations.}
\footnotesize
\begin{tabular}{cccccccccc}

\hline 

 Planet & RA & DEC & Sp Typ &Age&d& $M_P$ ($M\sin i$)& $a$ & $S_\nu$ & $\nu_c$ \\ 
 name & hms  &    dms &     --& Gyr & pc & $M_J$ &au & mJy &~MHz \\
\hline 
\\
$\beta$ UMi b & 14 50 42.32  &      74 09 19.81 &     K4-III & 2.9 $\pm$ 1.0 & 38.8 $\pm$ 0.8  & 6.1 $\pm$ 1 & 1.4 $\pm$ 0.1 & {0.1} & 478\\ 
\\
\hline
\\
$\alpha$~Tau b &04 35 55.20 &16 30 35.10&     K5-III &6.6  $\pm$ 2.4& 20.4 $\pm$ 0.3 &6.5 $\pm$ 0.53 & 1.46 $\pm$ 0.27 & {0.6} & 528 \\ 
\\
\hline 
\\
$\beta$ Gem b &07 45 18.94 &28 01 34.31&     K0-III &0.7 & 10.4 $\pm$ 1 &2.9 $\pm$ 0.1 & 1.7 $\pm$ 0.03 & {15.3} & 139 \\ 
\\
\hline 
\end{tabular} 
\label{Table1er}

\end{center}

\end{table*}

\subsection{$\beta$~UMi~b}
$\beta$~UMi (Kochab) is a 2.9 $\pm$ 1.0 Gyr old, K4-III giant at 38.8 $\pm$ 0.8 pc from Earth \citep{Lee14}. The system hosts a super-Jupiter $\beta$~UMi~b with mass ($M\sin i$) of 6.1 $\pm$ 1 $M_J$ \citep{Lee14}. The planet $\beta$~UMi~b is at 1.4 $\pm$ 0.1 au from the host star. The maximum cyclotron frequency $\nu_c$ for radio emission from the planet (based on Equation \ref{nuer}) is 478~MHz. { Based on our calculations, the expected flux density from the planet is 0.1 mJy.} $\beta$~UMi~b was previously observed by \cite{gorman18} with Low-Frequency Array (LOFAR) at 150~MHz. At 150~MHz, no emission was detected from the source, and a 3$\sigma$ upper limit of 0.57 mJy was placed for the emission at 150~MHz. 

\subsection{$\alpha$~Tau~b}
$\alpha$~Tau~b is a 6.5 $\pm$ 0.53 $M_J$ mass ($M\sin i$) planet at a distance of 1.46 $\pm$ 0.27 au from the host star \citep{Hatzes15}. The host star $\alpha$~Tau  (Aldebaran) is a 6.6 Gyr $\pm$ 2.4 old K5-III giant at 20.4 $\pm$ 0.3 pc from Earth. Using Equation \ref{nuer}, we find that $\nu_c$ for $\alpha$~Tau~b is 528~MHz. This falls in the middle of band~3 (250-500~MHz) and band~4 (550-900~MHz) of uGMRT.{ The expected flux density from  $\alpha$~Tau~b  is 0.6 mJy.} The host star $\alpha$~Tau has been observed at higher frequencies ($\geq$ 3.15 GHz) by \cite{Gorman13}. At 3.15 GHz, the integrated flux density was 0.04 mJy. {The high-frequency emission is due to thermal free-free interactions in the stellar atmosphere and not due to the star-planet interaction.} No low-frequency observations of the system have been carried out. 

\subsection{$\beta$~Gem~b}
The third system we analyzed in this work is $\beta$ Gem (Pollux). $\beta$ Gem is a 0.7 Gyr old K0-III giant \citep{Takeda08}. The star is host to a 2.9 $\pm\, 0.1 \, M_J$ mass ($M\sin i$) planet at 1.7 au \citep{Reffert06}. The host star is also relatively close to Earth at 10.4 $\pm$ 1 pc. B{ased on the planet's mass, the maximum emission frequency is 138.6~MHz, close to band~2 (150~MHz) of GMRT with the expected flux density of 15.3 mJy.}  The system was previously observed with LOFAR at 150~MHz by \cite{gorman18}. At 150~MHz using LOFAR, the rms sensitivity of 325 $\mu$Jy/beam was attained, however, no radio emission from the system was detected.     

\section{Observation and data reduction}

The uGMRT observations for $\alpha$~Tau and $\beta$~UMi were carried out on 2020 February 15 and 2020 February 16  as part of the proposal number  $37\_006$ (PI-Mayank Narang). The sources $\beta$~UMi~b and $\alpha$~Tau~b were observed in band~3 (250-500~MHz) and band~4 (550-900~MHz) of uGMRT. In band~3, we observed at 400~MHz with a bandwidth of 200~MHz, while in band~4, we observed at 650~MHz with a bandwidth of 200~MHz. 

In band~3 at 400~MHz (with a bandwidth of 200~MHz), \UMi~was observed for 4 hrs with 3C286 as the band-pass and flux calibrators and 1400+621 as the phase calibrator. The band-pass and flux calibrators were observed twice, once at the beginning of the observation and once at the end of the observation, while the phase calibrator was observed regularly during the observations in a loop of 20 minutes on sources and 5 minutes on the phase calibrator. At 650~MHz (with a 200~MHz bandwidth), \UMi was observed for 3 hrs in band~4 of uGMRT. The band-pass and flux density calibrators used were 3C286 and 3C147 and were observed twice, once at the beginning (3C147) and once at the end of the observation (3C286). The phase calibrator used was 1400+621 and was observed in a loop of 20 minutes on \UMi and 5 minutes on the phase calibrator. 

In band~3, \alfT was observed at 400~MHz (bandwidth 200~MHz) for 5 hrs with the band-pass and flux calibrators being 3C48 and 0521+166 being the phase calibrator. 3C48 was observed twice, once at the beginning and once at the end of the observation, while the phase calibrator was observed in a loop with 20 minutes on \alfT and 5 minutes on 0521+166. \alfT was also observed for 4 hrs in band~4 (650~MHz and bandwidth 200~MHz) of uGMRT. Again, 3C147 and  3C48 were used as the band-pass and flux calibrators, and 0521+166 was used as the phase calibrator. 3C48 was observed at the beginning of the observation, while 3C147 was observed at the end. The phase calibrator 0521+166, similar to the previous observations, was observed in a loop with 20 minutes on the science target $\alpha$~Tau and 5 minutes on the phase calibrator (0521+166).

\begin{figure*}
\centering
\includegraphics[width=0.4\linewidth]{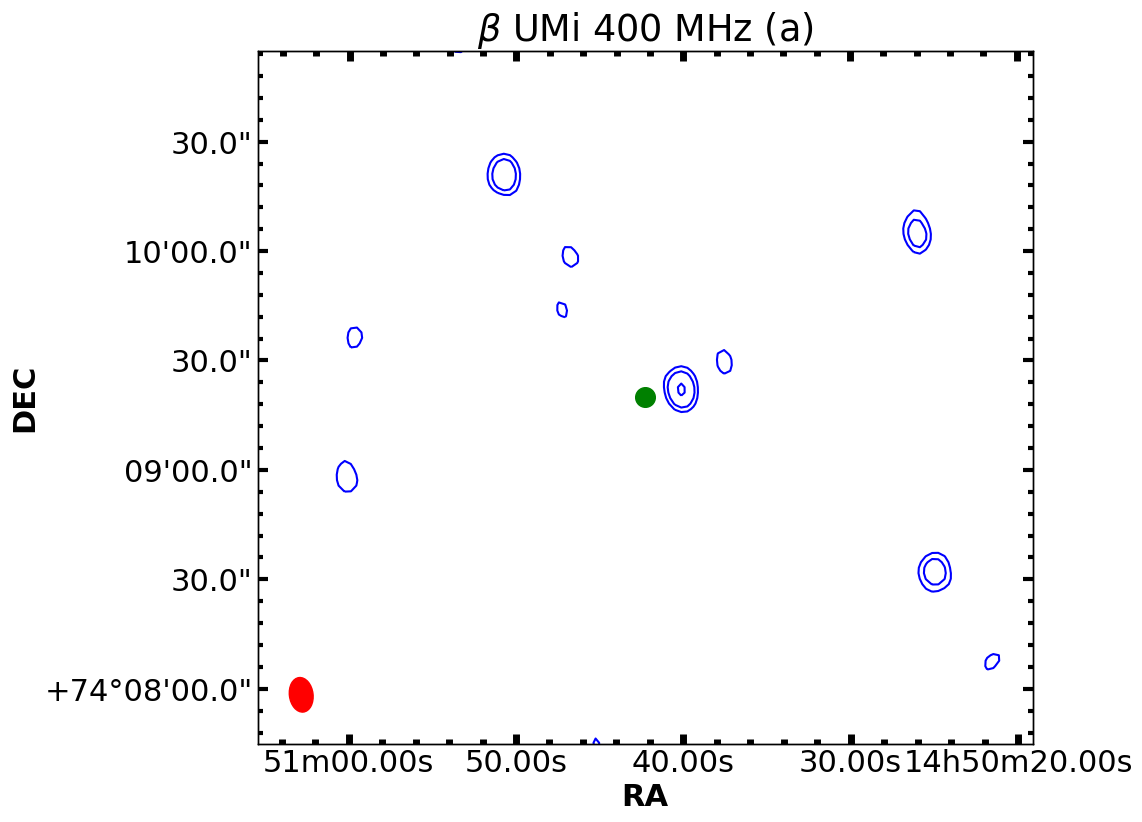}\includegraphics[width=0.4\linewidth]{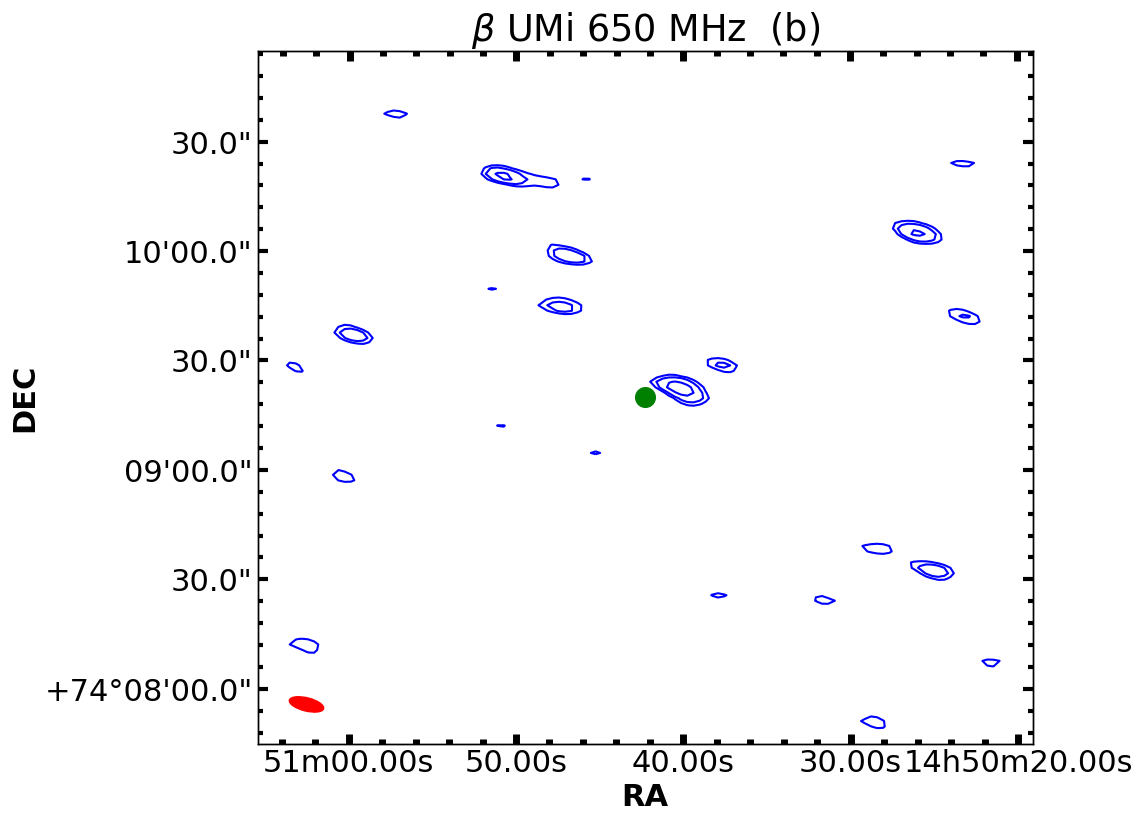}
\caption{{The uGMRT image (blue contours) of the \UMi~field at 400~MHz (a) and 650~MHz (b).  The green circle marks the position of the planetary system after correcting for proper motion using Hipparcos data \citep{2007A&A...474..653V}. The contours plotted are  5, 10, 30, and 50~$\times\;\sigma$. The beam is shown as a red ellipse at the bottom left corner. } }
\label{fig2_er}
\end{figure*}

We also retrieved archival legacy GMRT observations for the $\beta$ Gem system. $\beta$ Gem was observed at 150~MHz with GMRT at two epochs {(Proposal number 24\_013 and 25\_039; PI-Eamon Gerard O'Gorman)}. The first observation was carried out on 2013 June 02 for 11 hrs. The band-pass and flux calibrator used were 3C48 and 3C286. 3C48 was observed for 10 minutes at the beginning of the observation, while 3C286 was observed for 9 minutes at the end. The phase calibrator used was 0744+378 and was observed in a loop with 52 minutes of the science target $\beta$ Gem and 9 minutes on the phase calibrator. The second observation was carried out on 2013 November 22. The observation setup was the same as the previous observation. In table \ref{Table2_er}, we summarize the observational setup.    

We reduced these uGMRT images using the  CASA Pipeline-cum-Toolkit for Upgraded Giant Metrewave Radio Telescope data REduction uGMRT- (CAPTURE) pipeline \citep{CAPTURE}. The reduced images were further corrected for falling sensitivity at the beam edges using the CASA task \textit{wbpbgmrt}\footnote{https://github.com/ruta-k/uGMRTprimarybeam} to produce the final images. For the legacy GMRT observations of $\beta$ Gem at 150~MHz, we used the Source Peeling and Atmospheric Modeling (SPAM) pipeline \citep{Intema09b, Intema14b, Intema14}.

\begin{table*}
\centering
\caption{{The observation summary, the rms sensitivity reached, and the beam sized for our observations are listed in this table.}}
\footnotesize
\begin{tabular}{ccccccccccc}
\hline

Source & band~& Date of  & Central  & Bandwidth & Obs time & Flux density & Phase  & rms& bmaj & bmin \\
 &  & observation  & frequency &  &  & calibrator & calibrator  & \\
-- & -- & -- &~MHz &~MHz & hrs & -- & -- & $\mu$Jy & $^{\prime\prime}$  & $^{\prime\prime}$\\ 

\hline
\\
\UMi & 3 & 2020 Feb 16 & 400 & 200 & 4 & 3C286 & 1400+621 & 25.8 & 9.5 &5.8\\
\\
\hline
\\
\UMi & 4 & 2020 Feb 16 & 650 & 200 & 3 & 3C286 $\&$ 3C147 & 1400+621 & 11.2 & 8.7 &3.6 \\
\\
\hline
\\
\alfT & 3 & 2020 Feb 15 & 400 & 200 & 5 & 3C48 & 0521+166 & 29 &6.9 &5.3 \\
\\
\hline
\\
\alfT & 4 & 2020 Feb 15 & 650 & 200 & 3 & 3C147 $\&$ 3C48 & 0521+166 & 9.4 &5.1 &3.9
\\
\\
\hline
\\
$\beta$ Gem b & 2 & 2013 June 02 & 150 & 6 & 11 & 3C286 $\&$ 3C48 & 0744+378 & 2900 & 20.8 & 16
\\
\\
\hline
\\
$\beta$ Gem b & 2 & 2013 Nov 22 & 150 & 6 & 11 & 3C286 $\&$ 3C48 & 0744+378 & 850 &24.3 &18
\\
\\
\hline
\end{tabular}

\label{Table2_er}
\end{table*}

\section{Results}
No radio emission was detected at the position of \UMi at 400~MHz or 650~MHz (see Figure \ref{fig2_er}). At 400~MHz with 4 hrs of integration time, we achieved an rms of 25.8 $\mu$Jy/beam corresponding to a 3$\sigma$ {upper limit} of 77.4 $\mu$Jy/beam for \UMi. In band~4 at 650~MHz, we achieved an rms of 11.2 $\mu$Jy/beam, which corresponds to a 3$\sigma$ {upper limit} of 33.6 $\mu$Jy/beam. A radio source close to the system (9.36$^{\prime\prime}$ away from $\beta$~UMi) was, however, detected both at 400~MHz and 650~MHz (see Figure \ref{fig2_er}). The radio source detected near $\beta$ Umi has an integrated flux of 284 $\pm$ 31 $\mu$Jy at 400~MHz and  216 $\pm$ 20 $\mu$Jy at 650~MHz. The astrometric accuracy of the uGMRT images was determined by comparing the positions of uGMRT-detected sources with those from Very Large Array Sky Survey (VLASS) \citep{2021ApJS..255...30G}. The positional offset between the VLASS position and the uGMRT position is  $\leq$~2$^{\prime\prime}$.  Given the astrometric accuracy of uGMRT, it is likely that the emission is not from the system.    

\begin{figure*}
\centering
\includegraphics[width=0.4\linewidth]{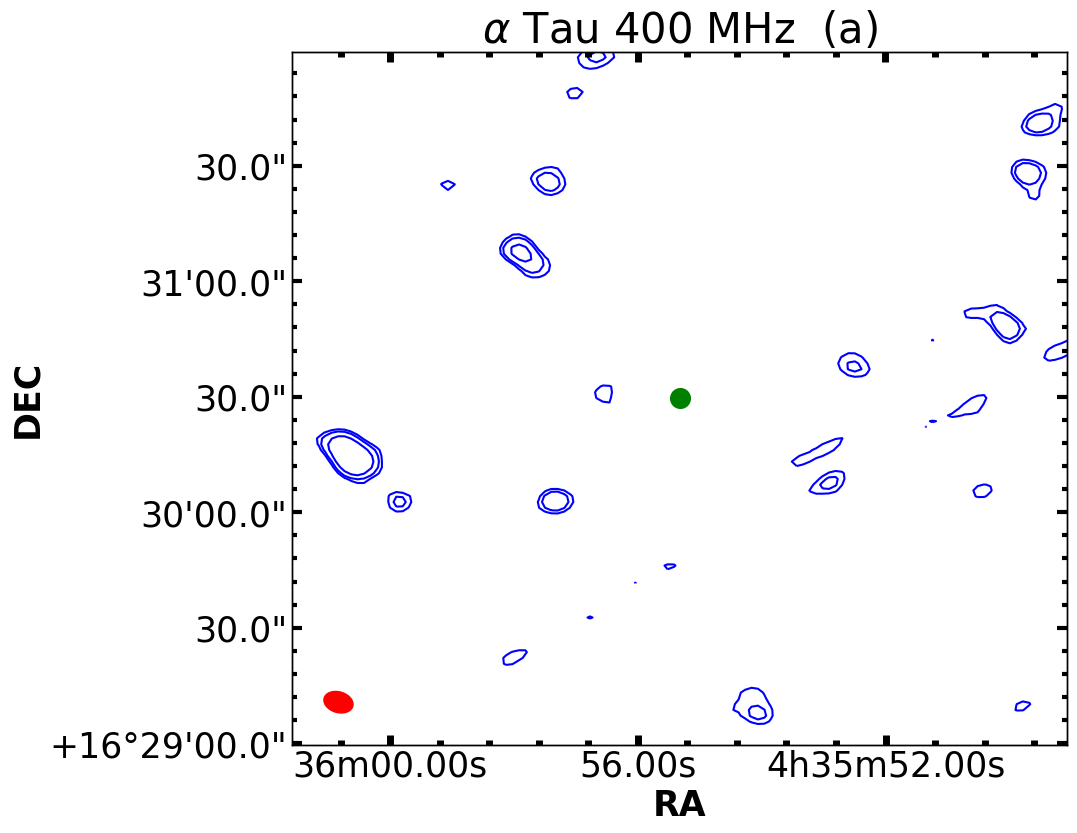}\includegraphics[width=0.4\linewidth]{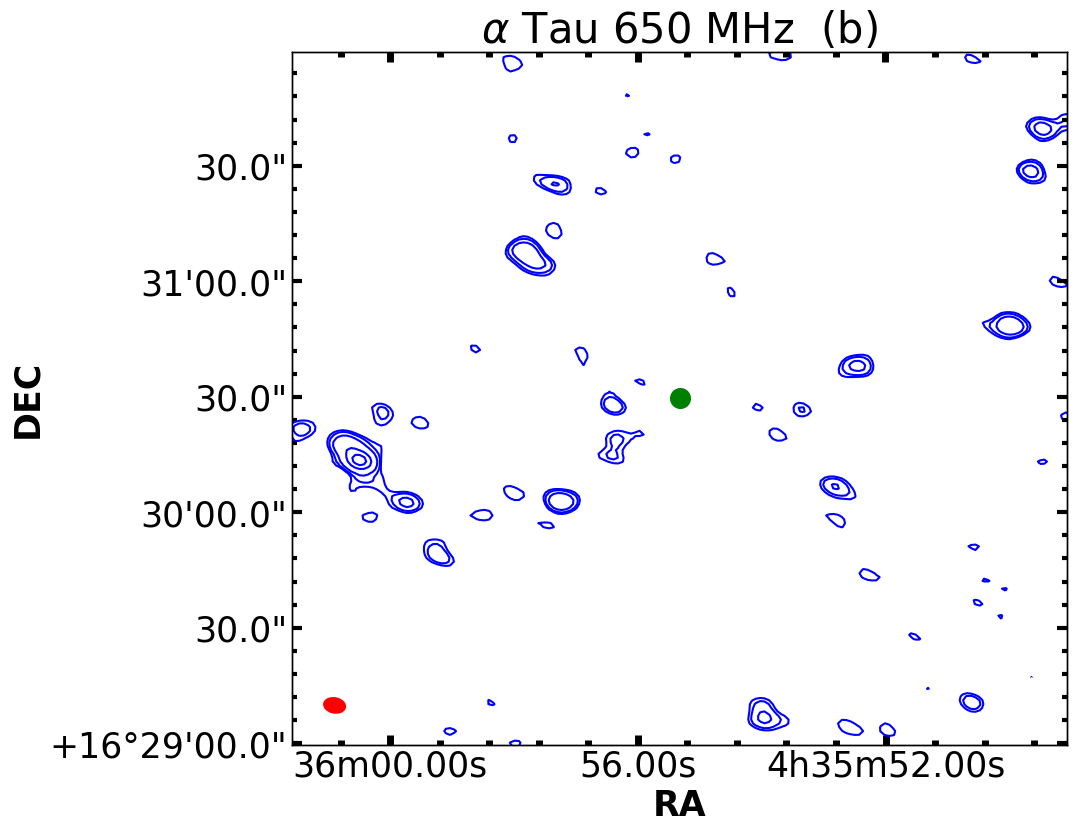}
\caption{{The uGMRT image (blue contours) of the \alfT~field at 400~MHz (a) and 650~MHz (b). The green circle marks the position of the planetary system after correcting for proper motion using Hipparcos data \citep{2007A&A...474..653V}. The contours plotted are 3, 5, 10, 30 and 
 50 $\times\;\sigma$. The beam is shown as a red ellipse at the bottom left corner.} }
\label{fig3_er}
\end{figure*}

\begin{figure*}
\centering
\includegraphics[width=0.4\linewidth]{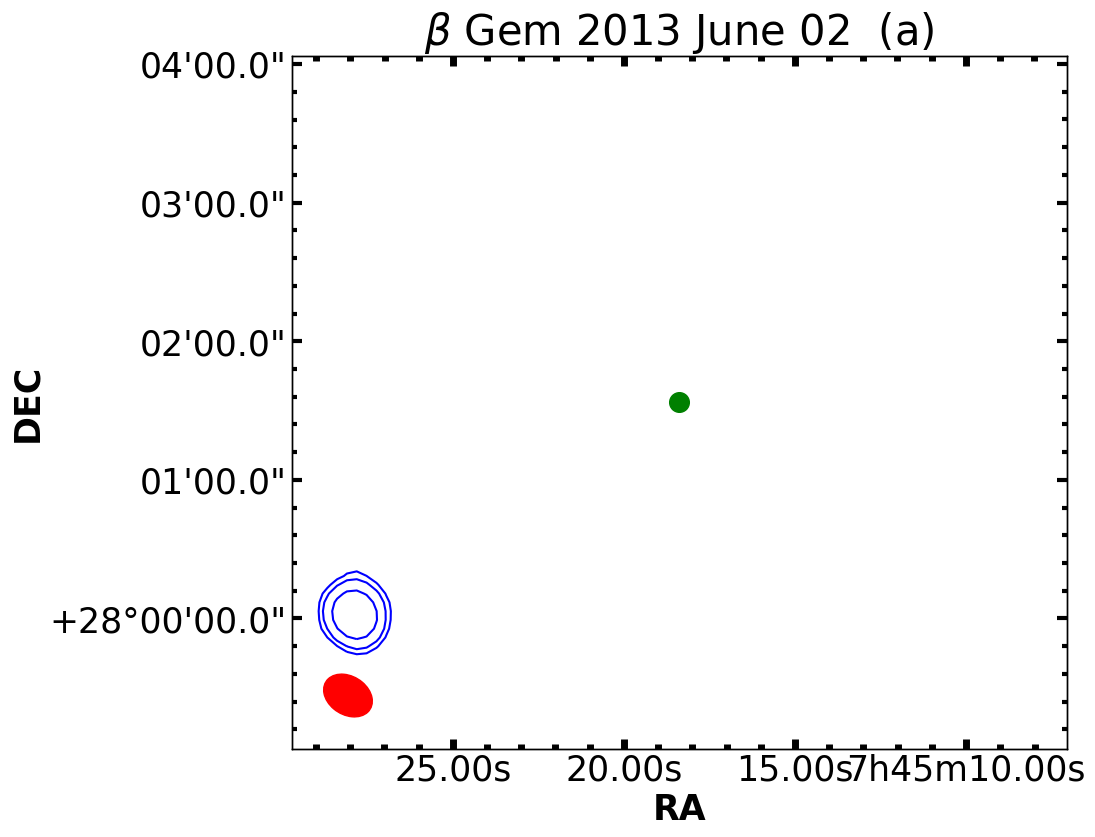}\includegraphics[width=0.4\linewidth]{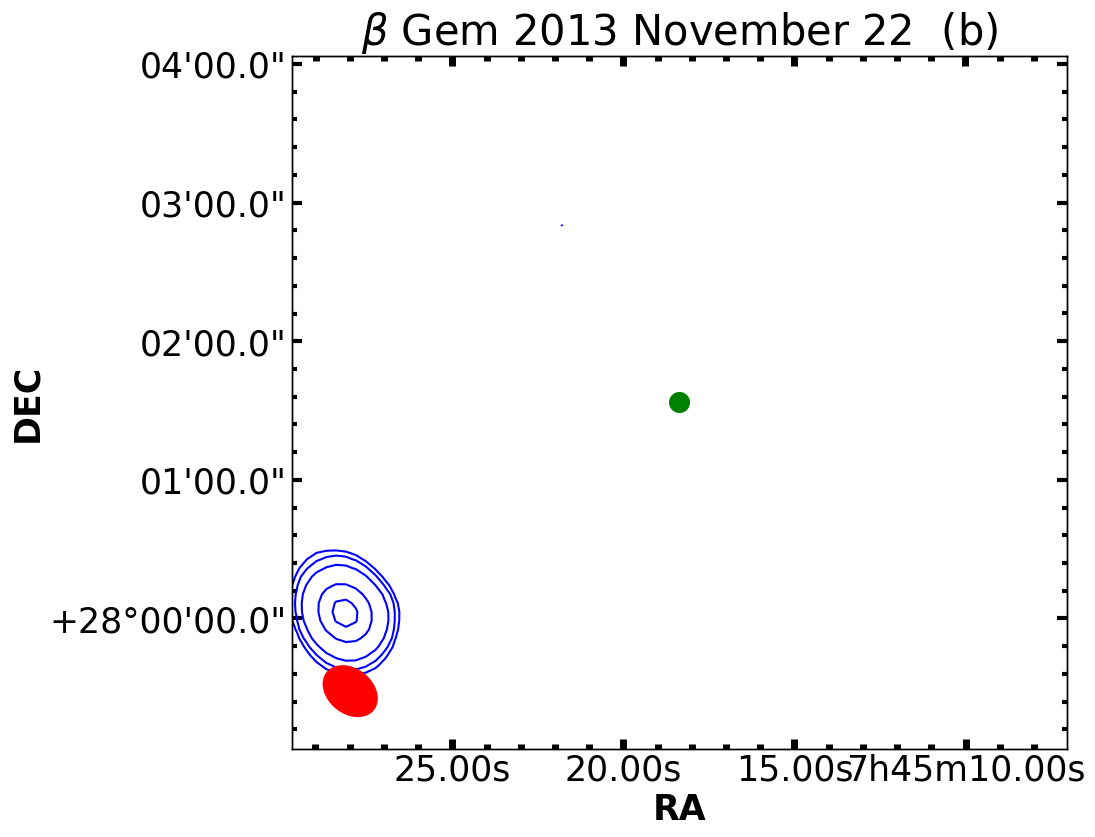}
\caption{{The GMRT image (blue contours) of the $\beta$ Gem field at 150~MHz. The green circle marks the position of the planetary system after correcting for proper motion using Hipparcos data \citep{2007A&A...474..653V}. The contours plotted are 3, 5, 10, and 30 $\times\;\sigma$. The beam is shown as a red ellipse at the bottom left corner.} }
\label{fig4_er}
\end{figure*}

In Figure \ref{fig3_er}, we show the 400~MHz and 650~MHz uGMRT images of the $\alpha$~Tau field. We do not detect any {emission} from the system. We obtained  an rms value of 29 $\mu$Jy/beam at 400~MHz (3$\sigma$ = 87 $\mu$Jy/beam)  and an rms value of 9.4 $\mu$Jy/beam (3$\sigma$ = 28.2 $\mu$Jy/beam)  at 650~MHz. These rms values are similar to those obtained for the $\beta$ UMi field. The 150~MHz observations of $\beta$ Gem are shown in Figure~\ref{fig4_er}. We do not detect any emission from the system at the two epochs. We were, however, able to obtain an rms value of 2.9 mJy/beam on 2013 June 02 and 0.85 mJy/beam on 2013 November 22.  \cite{gorman18} also observed the source at 150~MHz with LOFAR. They archived an rms of 0.325 mJy/beam, a factor of 2 smaller than what we achieved for the system using GMRT.

\section{Discussion}

\begin{figure*}
\centering 
\includegraphics[width=1\linewidth]{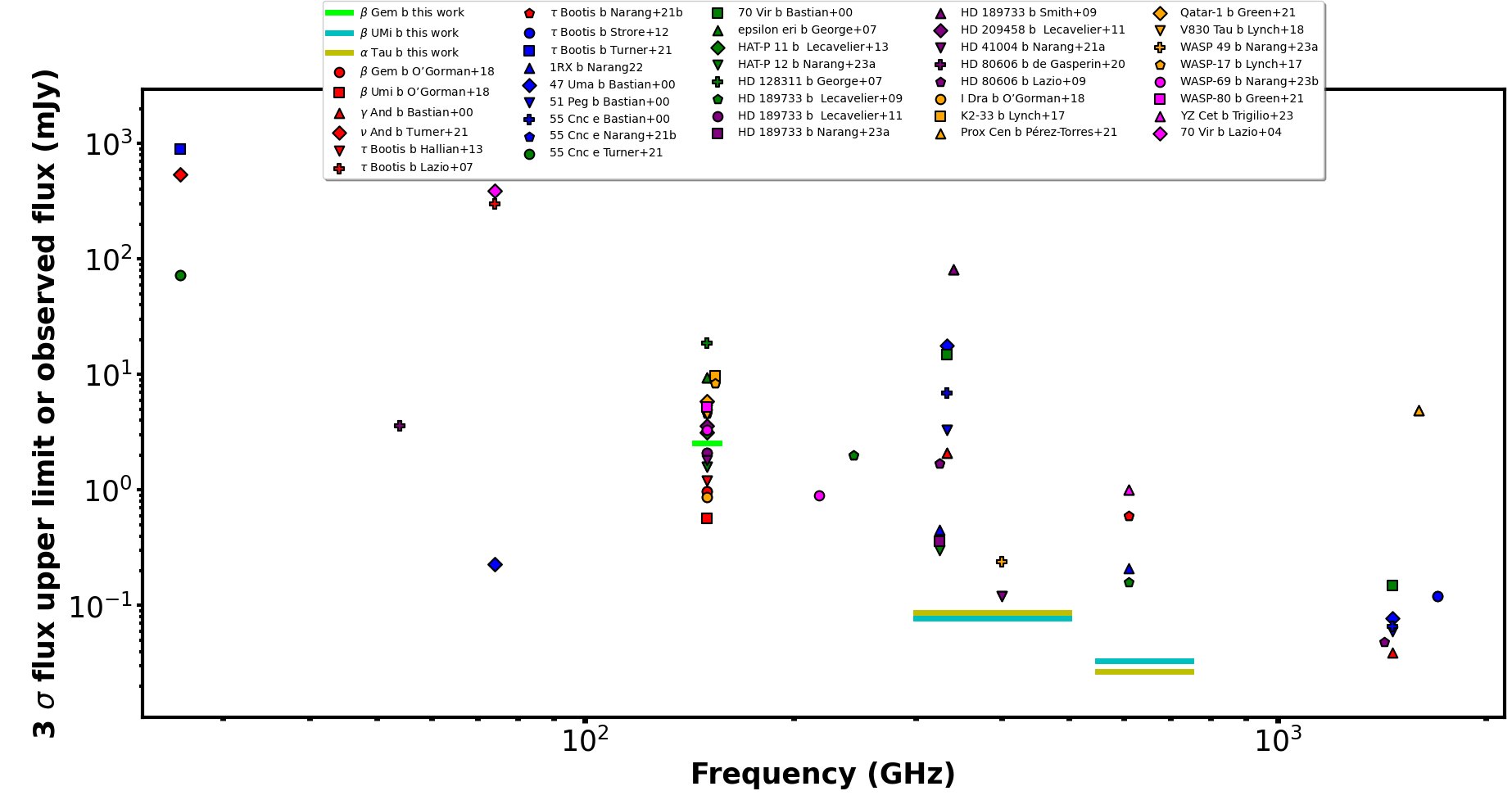}
\caption{The upper limits to the flux density {for} some of the previous attempts at detecting radio emission from exoplanets as a function of the frequency of observations \citep{gorman18, Bastian00,Turner20,Hallinan13, Lazio07, Narang21b, Narang21, Stroe12, 2022MNRAS.515.2015N, Lazio04, Lynch18,  George07, Etangs13, Etangs11, Etangs09,smith09, 2020A&A...644A.157D, 2021MNRAS.500..211G, 2023AJ....165....1N, 2023MNRAS.522.1662N, Route2019,Lynch17} along with the 3 $\sigma$ upper limits to the flux density we reached in this work (solid line) for the 150~MHz, 400~MHz and 650~MHz uGMRT observations. Also shown are the detections of radio emission due to possible Star-planet interaction from the M dwarfs Proxima Centauri \citep{prez20} and YZ Cet \citep{2023arXiv230500809T} as well as the tentative detection from $\tau$ Bootis b \citep{Turner20}.}
\label{fig5}
\end{figure*}

{The rms sensitivities that we reached are the deepest achieved at these frequencies for an exoplanet field by a factor of 1.6 at 400~MHz \citep[see][]{Narang21} and a factor of 5.7 at 650~MHz \citep[see][]{Etangs09}.} In Figure \ref{fig5}, we compare the upper limits to the flux density of observations that have been carried out to detect radio emission from exoplanets with those achieved in this work.  
The 3$\sigma$ upper limits that we derive for the radio flux density from \UMi,  \alfT $\beta$ Gem are smaller than the predicted flux density. In the literature, there have been several reasons proposed for why no radio emission has been detected from exoplanets \citep[e.g.,][]{Bastian00, Jardine08, Hallinan13, Narang21}. However, most of the reasons discussed only apply to hot Jupiters, as described in Section 1. In the following subsection, we discuss possible reasons why no radio emission was detected from these three systems of planet-bearing evolved stars.

\subsection{Overestimation of flux density}
{One of the major sources of uncertainty in the flux density estimation is the mass loss rate from the host star and $V_\infty$. To have a uniform estimate for radio emission from all planets around evolved stars, we used an analytical prescription for mass loss rate ($\dot{M}$) as given by Equation \ref{mass_losseq} and followed the prescription for stellar wind ionization fraction and  $V_\infty$ from \cite{1986AJ.....91..602D}.} The actual $\dot{M}$ from the host star might be different. The mass loss from $\alpha$~Tau is measured to be $1.6-6.5 \times 10^{-11} M_\odot$/yr \citep{Judge91,Gorman13} while from Equation \ref{mass_losseq} the mass loss{$\dot{M}$ is  $4.6 \times 10^{-10} M_\odot$/yr.} The new expected flux density would be between  42-130 $\mu$Jy (assuming $V_\infty=30\, km/s$).   Similarly \cite{Judge91} and \cite{Cranmer11} have computed the $\dot{M}$ from $\beta$ UMi to be $\sim$ $1 \times 10^{-10} M_\odot$/yr while our predictions estimate the mass loss rate {$\dot M$ to be  $2.2 \times 10^{-10} M_\odot$/yr.} The predicted flux density based on the observed mass-loss is 55 $\mu$Jy (using $V_\infty=30~km/s$). Given the sensitivity of our observations and the expected flux density from \alfT and \UMi, we still could have detected the emission at a  2$\sigma$ level.


\subsection{Overestimation of cyclotron frequency}
{The maximum cyclotron frequencies for \UMi and \alfT using Equation \ref{nuer} are 478~MHz and 528~MHz, respectively; hence these sources should be observed in band~3 and band~4 of uGMRT.} For $\beta$ Gem, the maximum cyclotron frequency was 138.6~MHz, close to band~2 (150~MHz) in legacy GMRT. 
However, several assumptions were made while deriving the maximum cyclotron frequency (Eq \ref{nuer}). \cite{Turner20} detected tentative radio emission from the $\tau$ Bo\"otis system between 14-21~MHz. $\tau$ Bo\"otis~b has an $M_{P}$ of $\sim 6 M_J$. This is similar to the mass of \UMi~and \alfT.  The maximum cyclotron frequency of these exoplanets may also be of the order of a few tens of~MHz. Further observations at lower frequencies are required to rule out this hypothesis.

\subsection{Variable/Episodic emission}
The low-frequency radio emission from the solar system planets is highly time variable.  It is possible that even in these systems, the radio emission might be time-variable. The recent radio emission (attributed to Star Planet Interaction SPI) detected from GJ~1151 \citep{Vedantham20} and Proxima Centauri system \citep{prez20} are also variable. Furthermore, radio emission from isolated brown dwarfs is also time-variable \citep[e.g.,][]{2007ApJ...663L..25H,2017ApJ...834..117W,2020ApJ...903...74R}. Hence, multiple observations that cover the complete orbital phase might help us rule out variability as a reason for the non-detections. However, these kinds of observations for long-period planets are not possible.

\subsection{Beamed Emission}
The Jovian decameter emission is only detectable over certain rotational phase ranges of the planet. This is because the emission is narrowly beamed. If we assume that the planet's magnetic axis is aligned with the planet's rotation axis and that emission is only produced near the planet's poles, then the beaming angle is between 50\textdegree $-\,$ 60\textdegree \citep{Melrose82, Dulk85, Zarka98, Treumann06}. The same may be true for these giant planets. Hence, more observations of these systems would be required to rule out beaming as a reason for the non-detections.   

\section{Summary}
In this work, we used the (upgraded) Giant Meterwave Radio Telescope (uGMRT) to search for radio {emission} from planets around evolved stars. We {argue} that the planets around evolved stars are better targets to observe radio emission compared to hot Jupiters around main-sequence stars. Hot Jupiters are likely tidally locked and might not produce magnetic fields of adequate strength to produce detectable radio emission. Furthermore due to being so close to their host stars the magnetospheres of these planets might also be smaller leading to weaker emission.

With GMRT, we observed three exoplanetary systems, $\alpha$~Tau, $\beta$~UMi, and $\beta$~Gem. These three systems were carefully chosen based on our modeling, which showed that they would be detectable at GMRT frequencies. We do not detect any radio emission from the three systems but place strong 3$\sigma$ upper limits to the flux density from these objects. Even though we have not yet been able to detect radio {emission} from an exoplanet, we have produced some of the deepest images of exoplanet planet fields at low radio frequencies.  At 400~MHz, the 3$\sigma$ upper limit is 87 $\mu$Jy/beam for $\alpha$~Tau and 77.4 $\mu$Jy/beam for $\beta$~UMi. From our observations at 650~MHz, we place a 3$\sigma$ upper limit of 28.2 $\mu$Jy/beam for $\alpha$~Tau~b and 33.6 $\mu$Jy/beam for $\beta$~UMi~b. For $\beta$ Gem b, at 150~MHz, we place an upper limit of 2.5 mJy. The non-detections could be attributed to the emission not being as strong as predicted by the theory or the emission being variable. Hence, further observations at required to rule out radio emission from these systems.

\section{Data availability}

The data presented in this article are available on the GMRT archive. The GMRT data can be accessed from  https://naps.ncra.tifr.res.in/goa/,  with proposal id $37\_006$, $24\_013$, and $25\_039$. In addition, the reduced fits files are available on request from the corresponding author.

\section{Acknowledgement}

This work is based on observations made with the Giant Metrewave Radio Telescope, operated by the National Centre for Radio Astrophysics of the Tata Institute of Fundamental Research, and is located at Khodad, Maharashtra, India. We thank the GMRT staff for their efficient support of these observations. In addition, we acknowledge the support of the Department of Atomic Energy, Government of India, under Project Identification No. RTI4002. M.T. is supported by JSPS KAKENHI Grant Number JP18H05442. M.N. would like to thank Prof. Harish Vedantham at  ASTRON for his comments and suggestions. This research has used the NASA Exoplanet Archive, which is operated by the California Institute of Technology, under contract with the National Aeronautics and Space Administration under the Exoplanet Exploration Program. This research has also used NASA's Astrophysics Data System Abstract Service and the SIMBAD database, operated at CDS, Strasbourg, France.

\bibliographystyle{mnras}
\bibliography{exoplanet}
\bsp    
\label{lastpage}
\end{document}